\documentstyle[12pt,epsfig]{article}

\newcommand{\re}{\mathop{\mathrm{Re}}\nolimits}
\newcommand{\im}{\mathop{\mathrm{Im}}\nolimits}
\newcommand{\sign}{\mathop{\mathrm{sign}}\nolimits}
\newcommand{\arsinh}{\mathop{\mathrm{arsinh}}\nolimits}
\newcommand{\arcosh}{\mathop{\mathrm{arcosh}}\nolimits}

\catcode`@=11
\newcount\@tempcntc
\def\@citex[#1]#2{\if@filesw\immediate\write\@auxout{\string\citation{#2}}\fi
  \@tempcnta\z@\@tempcntb\m@ne\def\@citea{}\@cite{\@for\@citeb:=#2\do
    {\@ifundefined
       {b@\@citeb}{\@citeo\@tempcntb\m@ne\@citea\def\@citea{,}{\bf
?}\@warning
       {Citation `\@citeb' on page \thepage \space undefined}}%
    {\setbox\z@\hbox{\global\@tempcntc0\csname b@\@citeb\endcsname\relax}%
     \ifnum\@tempcntc=\z@ \@citeo\@tempcntb\m@ne
       \@citea\def\@citea{,}\hbox{\csname b@\@citeb\endcsname}%
     \else
      \advance\@tempcntb\@ne
      \ifnum\@tempcntb=\@tempcntc
      \else\advance\@tempcntb\m@ne\@citeo
      \@tempcnta\@tempcntc\@tempcntb\@tempcntc\fi\fi}}\@citeo}{#1}}
\def\@citeo{\ifnum\@tempcnta>\@tempcntb\else\@citea\def\@citea{,}%
  \ifnum\@tempcnta=\@tempcntb\the\@tempcnta\else
   {\advance\@tempcnta\@ne\ifnum\@tempcnta=\@tempcntb \else
\def\@citea{--}\fi
    \advance\@tempcnta\m@ne\the\@tempcnta\@citea\the\@tempcntb}\fi\fi}
\catcode`@=12

\begin{document}
\title{\vskip-3cm{\baselineskip14pt
\centerline{\normalsize DESY 00-079\hfill ISSN 0418-9833}
\centerline{\normalsize hep-ph/0007002\hfill}
\centerline{\normalsize July 2000\hfill}}
\vskip1.5cm
Higgs-Boson Production and Decay Close to Thresholds}
\author{{\sc Bernd A. Kniehl,$^1$ Caesar P. Palisoc,$^2$
Alberto Sirlin$^{1,3}$}\\
{\normalsize $^1$ II. Institut f\"ur Theoretische Physik, Universit\"at
Hamburg,}\\
{\normalsize Luruper Chaussee 149, 22761 Hamburg, Germany}\\
{\normalsize $^2$ National Institute of Physics, University of the
Philippines,}\\
{\normalsize Diliman, Quezon City 1101, Philippines}\\
{\normalsize $^3$ Department of Physics, New York University,}\\
{\normalsize 4 Washington Place, New York, New York 10003, USA}}

\date{}

\maketitle

\thispagestyle{empty}

\begin{abstract}
At one loop in the conventional on-mass-shell renormalization scheme, the
production and decay rates of the Higgs boson $H$ exhibit singularities 
proportional to $(2M_V-M)^{-1/2}$ as the Higgs-boson mass $M$ approaches from
below the pair-production threshold of a vector boson $V$ with mass $M_V$.
This problem is of phenomenological interest because the values $2M_W$ and
$2M_Z$, corresponding to the $W$- and $Z$-boson thresholds, lie within the $M$
range presently favoured by electroweak precision data.
We demonstrate how these threshold singularities are eliminated when the
definitions of mass and total decay width of the Higgs boson are based on the
complex-valued pole of its propagator.
We illustrate the phenomenological implications of this modification for the
partial width of the $H\to W^+W^-$ decay.

\medskip

\end{abstract}

\newpage

\section{Introduction}

The conventional definitions of the mass $M$ and the total decay width 
$\Gamma$ of an unstable boson are given by
\begin{eqnarray}
M^2&=&M_0^2+\re A(M^2),
\label{eq:mos}\\
M\Gamma&=&-\frac{\im A(M^2)}{1-\re A^\prime(M^2)},
\label{eq:gos}
\end{eqnarray}
where $M_0$ is the bare mass and $A(s)$ is the self-energy in the case of
scalar bosons and the transverse self-energy in the case of vector bosons.
In fact, most calculations of the total decay rates are based on 
Eq.~(\ref{eq:gos}).
$M$ and $\Gamma$ are conventionally referred to as the on-shell mass and width,
respectively.

However, over the last decade it has been shown that, in the context of gauge
theories, $M$ and $\Gamma$ become gauge dependent in $O(g^4)$ and  $O(g^6)$,
respectively, where $g$ is a generic gauge coupling \cite{sir,wil,kni}.
A solution to this problem can be achieved by defining the mass and width in
terms of the complex-valued position of the propagator's pole:
\begin{equation}
\bar s=M_0^2+A(\bar s),
\label{eq:sba}
\end{equation}
an idea that goes back to well-known postulates of scattering ($S$) matrix
theory \cite{pei}.
An important advantage of Eq.~(\ref{eq:sba}) is that $\bar s$ is gauge
independent to all orders in perturbation theory \cite{sir,wil,kni}.
A frequently employed parameterization is \cite{sir,con}
\begin{equation}
\bar s=m_2^2-im_2\Gamma_2,
\end{equation}
where we use the notation of Ref.~\cite{sir}.
Identifying $m_2$ and $\Gamma_2$ with the gauge-independent definitions of
mass and width, we have
\begin{eqnarray}
m_2^2&=&M_0^2+\re A(\bar s),
\label{eq:mpo}\\
m_2\Gamma_2&=&-\im A(\bar s).
\label{eq:gpo}
\end{eqnarray}
Alternative, gauge-independent definitions of mass and width based on
$\bar s$, with particular merits, have been discussed in the literature
\cite{sir,wil,kni,boh}.
A phenomenologically relevant application of the $S$-matrix approach is to
observables at the $Z$-boson resonance \cite{lei}.

Over a period of two decades, the on-shell renormalization scheme
\cite{oms,fle,kra} has provided a very convenient framework for the
calculation of quantum corrections in electroweak perturbation theory.
In fact, many important calculations have been performed in this scheme.
One of its principal aims is the parameterization of $S$-matrix elements in
terms of physical masses and coupling constants.
For most calculations at the one-loop level, Eqs.~(\ref{eq:mos}) and 
(\ref{eq:gos}) are satisfactory and, in fact, the original papers 
\cite{oms,fle} employed such definitions.
In higher orders, the gauge dependence of $M$ and $\Gamma$ precludes their
identification with physical quantities.
It is then natural to remedy this deficiency by replacing Eqs.~(\ref{eq:mos}) 
and (\ref{eq:gos}) by Eqs.~(\ref{eq:mpo}) and (\ref{eq:gpo}), respectively.
In this way, the calculations are parameterized in terms of constants, such as
$m_2$ and $\Gamma_2$, that can be identified with physical observables to all
orders in perturbation theory.
In particular, we observe from Eq.~(\ref{eq:mpo}) that the mass counterterm, a
basic quantity in the renormalization procedure, is given by
$\re A\left(\bar s\right)$, rather than $\re A(M^2)$.
We shall refer to this improved formulation, based on Eqs.~(\ref{eq:mpo}) and 
(\ref{eq:gpo}), as the pole scheme.

There is another significant pitfall of Eqs.~(\ref{eq:mos}) and 
(\ref{eq:gos}), which has gone almost unnoticed so far.
At the one-loop order, the production cross sections and total and partial
decay widths of the Higgs boson $H$ exhibit singularities proportional to
$(2M_V-M)^{-1/2}$ as the Higgs-boson mass $M$ approaches from below the
pair-production threshold of a vector boson $V$ with mass $M_V$
\cite{fle,bff,bvv,hww,hff,pr}.
This problem is of phenomenological interest because the values $2M_W$ and
$2M_Z$, corresponding to the $W$- and $Z$-boson thresholds, lie within the $M$
range presently favoured by electroweak precision data \cite{ewwg,nh}.
On the other hand, there is no such singularity at the pair-production
threshold of a fermion $f$ \cite{fle,bff,bvv,hww,hff,pr}.
This circumstance may be related, by the use of an appropriate dispersion
relation, to the different threshold behaviours of the lowest-order partial
widths of the decays $H\to VV$ and $H\to f\bar f$, which are proportional to
$(M-2M_V)^{1/2}$ and $(M-2M_f)^{3/2}$, respectively \cite{hww}.
In the case of a two-body threshold, this kind of singularity generally occurs
if the two particles form an $S$-wave state and the sum of their masses is
degenerate with that of the primary particle \cite{bha}.
For example, it would also occur in $Z$-boson production and decay if the
mass relation $M_Z=2M_t$, where $t$ denotes the top quark, were satisfied
\cite{jeg}, since in this case the $t\bar t$ pair can be in an $S$-wave state.
By the same token, it would occur for an extra neutral vector boson $Z^\prime$
with mass $M_{Z^\prime}=2M_t$.
For definiteness, in the following, we focus our attention on the Higgs boson
$H$ of the standard model (SM).

As explained in Refs.~\cite{hww,hff,pr}, the threshold singularity is an
artifact of treating an unstable particle, such as the Higgs boson, as an
asymptotic state of the $S$ matrix.
Detailed inspection \cite{hww,hff,pr} reveals that it originates from the
wave-function renormalization constant in the on-shell scheme,
\begin{equation}
Z=\frac{1}{1-\re A^\prime(M^2)},
\label{eq:zos}
\end{equation}
which also appears in the definition of $\Gamma$ in Eq.~(\ref{eq:gos}).
One way to obtain Eq.~(\ref{eq:zos}) is to consider the Taylor expansion of
the inverse propagator $s-M_0^2-A(s)$ about $s=M^2$.
This procedure tacitly assumes that $A(s)$ is analytic near $s=M^2$, so that
the Taylor expansion can be performed.
In most cases, this assumption is valid.
However, $A(s)$ possesses a branch point if $s$ is at a threshold.
If the threshold is due to a two-particle state with zero orbital angular
momentum, then $\re A^\prime(s)$ diverges as $1/\beta$, where $\beta$
is the relative velocity common to the two particles, as the threshold is
approached from below \cite{bha}.
Another case where the non-analyticity of $A(s)$ in the neighbourhood of 
$s=M^2$ leads to serious problems in the on-shell formalism is the behaviour
of the resonant amplitude when the unstable particle is coupled to massless 
quanta, such as photons and gluons \cite{pas}.

The purpose of this paper is to show how the threshold singularities are 
rigorously removed by adopting the pole scheme.
The salient point is that $Z$ is redefined in such a way that the derivative
term $\re A^\prime(M^2)$ is replaced by an appropriate ratio of differences,
where the increment of the argument is of order mass times width.
In this way, the threshold singularities are regularized by the very width of
the primary particle.
This mechanism was illustrated for a toy model, consisting of a real scalar
particle coupled to two stable, complex scalar particles, through a numerical
simulation in Ref.~\cite{bha}.
Here, we analytically elaborate the underlying formalism in a general,
model-independent way and apply it to a case of phenomenological interest,
namely the thresholds of the SM Higgs boson at $M=2M_V$, with $V=W,Z$.

If the threshold particles are unstable, an alternative way of eliminating
the threshold singularities in a physically meaningful way is to incorporate
their widths in the one-loop calculation \cite{mel}.
Notice that the on-shell and pole formulations are equivalent through the 
one-loop order, as may be seen, for example, by Taylor expanding 
Eq.~(\ref{eq:gpo}) about $m_2^2$ \cite{sir,kni}.
Furthermore, as will become apparent later on, threshold singularities only
appear in connection with physical thresholds.
Therefore, this regularization procedure does not spoil the gauge
independence of the physical predictions.
If the widths of the threshold particles are much larger than the one of the
primary particle, then it appears plausible to adopt this method.
In general, however, the two regularizing effects, associated with the widths
of the primary and threshold particles, should be combined in a unified
analysis.
We explain how this can be achieved.

This paper is organized as follows.
In Section~\ref{sec:os}, we present the one-loop expressions for the 
Higgs-boson self-energy in $R_\xi$ gauge \cite{rxi} and in the
pinch-technique (PT) framework \cite{cor,deg}, explicitly exhibit the
threshold singularities at $M=2M_V$, and show that the latter are gauge 
independent.
In Section~\ref{sec:pol}, starting from the definition of $\Gamma_2$ in
Eq.~(\ref{eq:gpo}), we derive the counterpart of Eq.~(\ref{eq:zos}) in the
pole scheme and show that it is devoid of threshold singularities.
We also present a simple substitution rule which allows us to translate
existing results for Higgs-boson observables from the on-shell scheme to the
pole scheme if the threshold particles are stable. 
In Section~\ref{sec:uns}, we generalize this substitution rule to the general
case of unstable threshold particles and discuss in detail the situation when
the widths of the latter are much larger than the one of the primary particle.
In Section~\ref{sec:dis}, we present a numerical analysis for the
$H\to W^+W^-$ partial decay widths of the Higgs boson, based on the results
from Ref.~\cite{hww}.
In Section~\ref{sec:con}, we summarize our conclusions.

\section{On-shell formulation
\label{sec:os}}

In this section, we pin down the origin of the threshold singularities in
Higgs-boson observables at $M=2M_V$ by inspecting the corresponding expression
of $Z$ in Eq.~(\ref{eq:zos}).
The relevant ingredient is the Higgs-boson self-energy $A(s)$.
Since we wish to clarify if the threshold singularities give rise to spurious
gauge dependence, we consider the one-loop expressions for $A(s)$ in $R_\xi$
gauge \cite{rxi} and in the PT framework \cite{cor,deg}.

\begin{figure}[ht]
\begin{center}
\centerline{\epsfig{figure=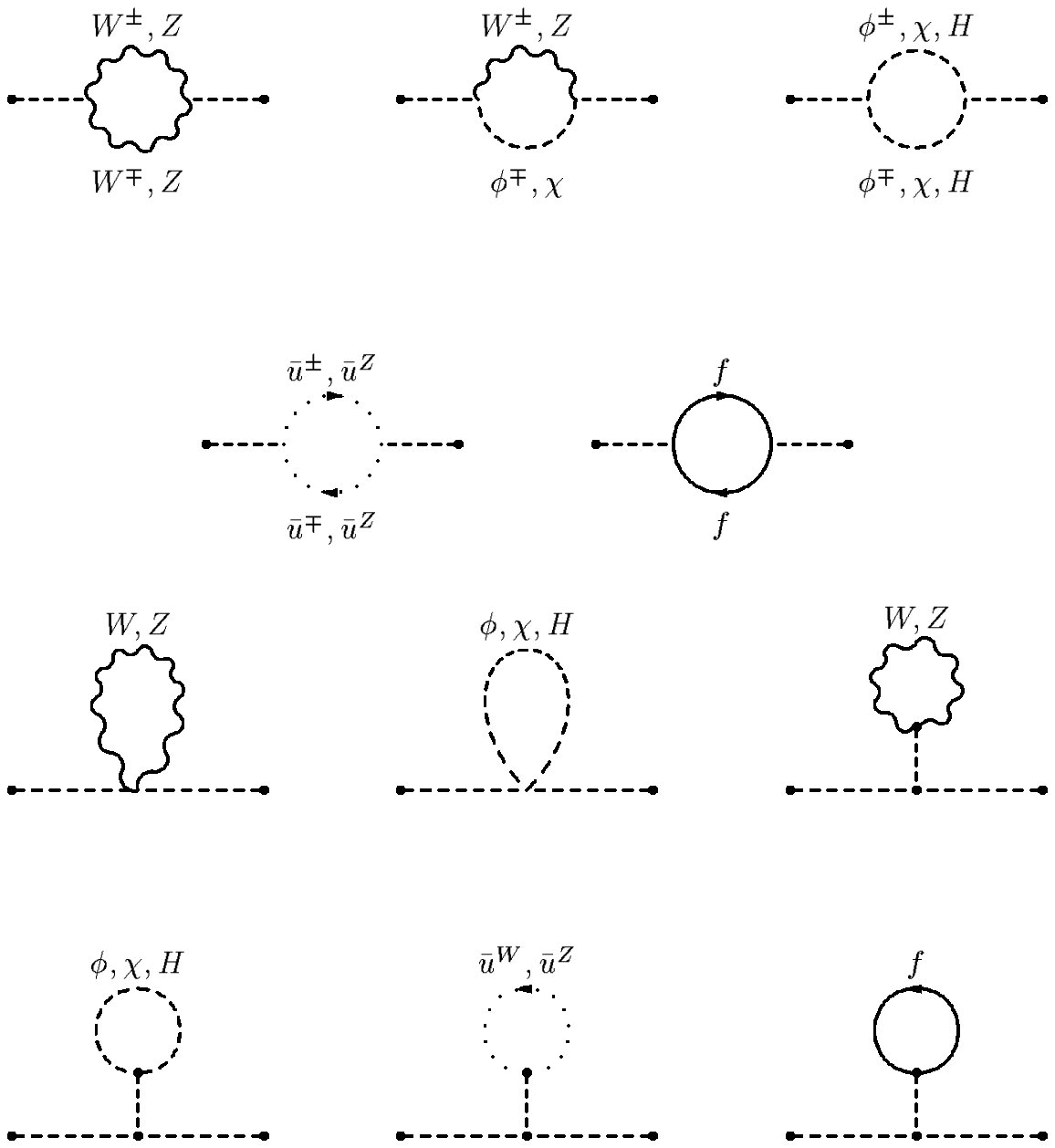,width=16cm,bbllx=131pt,bblly=372pt,%
bburx=466pt,bbury=735pt,clip=}}
\caption{Feynman diagrams pertinent to the conventional self-energy of the SM
Higgs boson in $R_\xi$ gauge.}
\label{fig:one}
\end{center}
\end{figure}

In $R_\xi$ gauge, we have to consider the Feynman diagrams depicted in
Fig.~\ref{fig:one}.
They yield
\begin{eqnarray}
A(s)&=&\frac{G}{\pi}\left\{-\left(\frac{s}{2}+3M_W^2\right)
A_0\left(M_W^2\right)
+\frac{1}{2}(s-M^2)A_0\left(\xi_WM_W^2\right)\right.
\nonumber\\
&&{}-\left(\frac{s^2}{4}-sM_W^2+3M_W^4\right)B_0\left(s,M_W^2,M_W^2\right)
\nonumber\\
&&{}+\frac{1}{4}(s^2-M^4)B_0\left(s,\xi_WM_W^2,\xi_WM_W^2\right)
+\frac{1}{2}(W\to Z)
\nonumber\\
&&{}-\frac{3}{4}M^2A_0(M^2)-\frac{9}{8}M^4B_0(s,M^2,M^2)
\nonumber\\
&&+\left.\sum_fN_fM_f^2\left[2A_0\left(M_f^2\right)
-\left(\frac{s}{2}-2M_f^2\right)B_0\left(s,M_f^2,M_f^2\right)\right]
\right\},
\label{eq:hxi}
\end{eqnarray}
where the sum is over fermion flavours $f$, $N_f=1$ (3) for leptons (quarks),
$G$ is related to Fermi's constant $G_\mu$ by
$G=G_\mu/\left(2\pi\sqrt2\right)$, $\xi_W$ is the gauge parameter associated
with the $W$ boson, and the term $(W\to Z)$ signifies the contribution
involving the $Z$ boson, which is obtained from the one involving the $W$
boson by replacing $M_W$ and $\xi_W$ with $M_Z$ and $\xi_Z$, respectively.
In dimensional regularization, the scalar one-loop one- and two-point 
integrals are defined as
\begin{eqnarray}
A_0\left(m_0^2\right)&=&-\frac{(2\pi\mu)^{4-D}}{i\pi^2}
\int\frac{d^Dq}{q^2-m_0^2+i\varepsilon},
\nonumber\\
B_0\left(p^2,m_0^2,m_1^2\right)&=&\frac{(2\pi\mu)^{4-D}}{i\pi^2}
\int\frac{d^Dq}{\left(q^2-m_0^2+i\varepsilon\right)
\left[(q+p)^2-m_1^2+i\varepsilon\right]},
\end{eqnarray}
where $\mu$ is the 't~Hooft mass scale and $D=4-2\epsilon$ is the space-time
dimensionality.
Their solutions in the physical limit $D\to4$ may, for example, be found in
Ref.~\cite{pr}.
The absorptive part of Eq.~(\ref{eq:hxi}) was already presented in 
Ref.~\cite{kni}.
In the 't~Hooft-Feynman gauge, with $\xi_W=\xi_Z=1$, Eq.~(\ref{eq:hxi})
agrees with the corresponding result given in Eqs.~(B.2) and (B.3) of
Ref.~\cite{hzz}.
Notice that $A(s)$ is gauge independent at $s=M^2$.
From Eqs.~(\ref{eq:mos}) and (\ref{eq:gos}) it hence follows that the one-loop
expression for $M$ and the tree-level one for $\Gamma$ are gauge independent,
too, as expected.
We note in passing that the gauge independence of $\re A(M^2)$ requires the
inclusion of the tadpole contribution in Eq.~(\ref{eq:hxi}).

\begin{figure}[ht]
\begin{center}
\centerline{\epsfig{figure=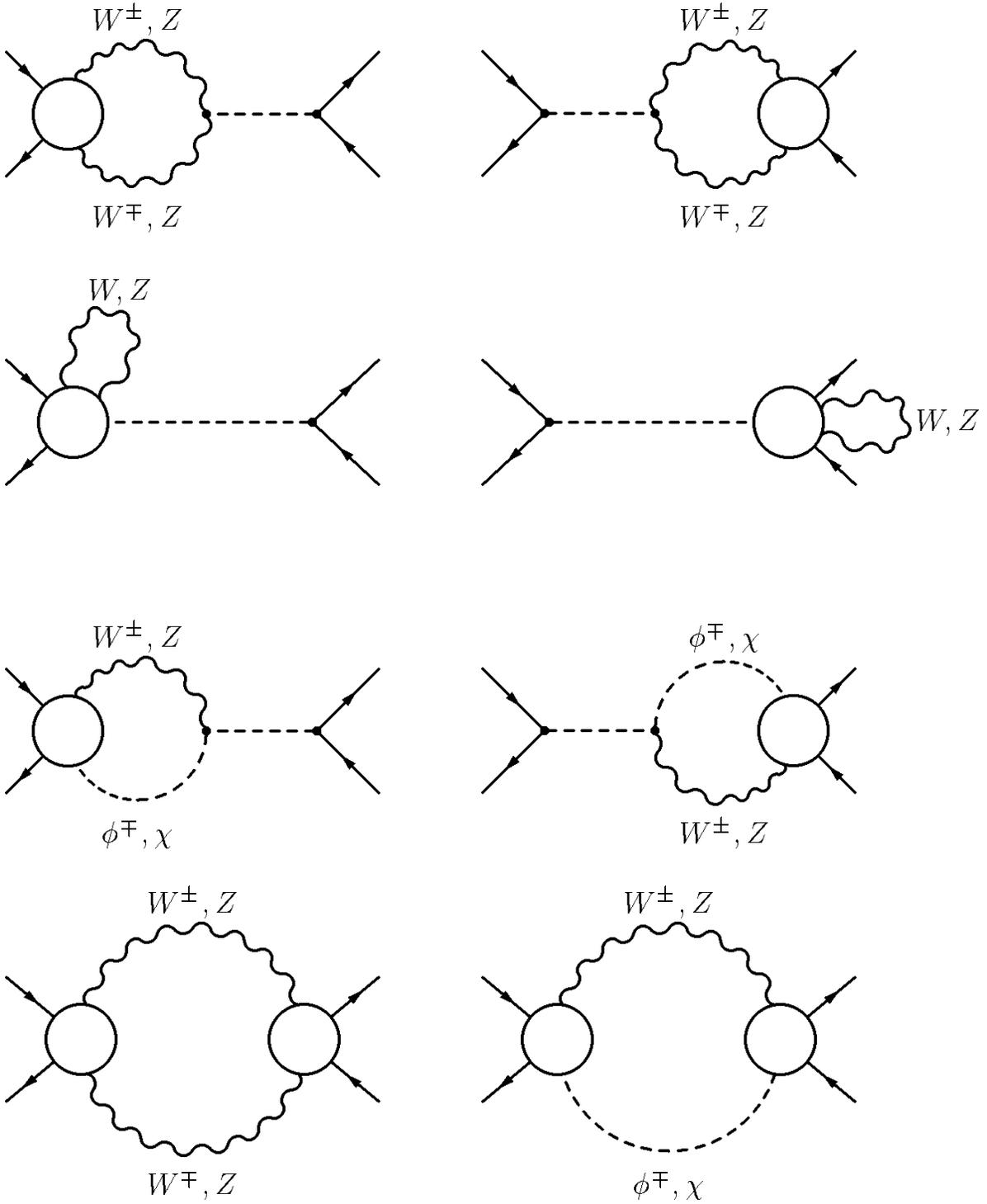,width=16cm,bbllx=158pt,bblly=351pt,%
bburx=478pt,bbury=742pt,clip=}}
\caption{Feynman diagrams pertinent to the pinch parts of the self-energy of
the SM Higgs boson in $R_\xi$ gauge.}
\label{fig:two}
\end{center}
\end{figure}

Next, we present the PT expression for $A(s)$.
We recall that the PT is a prescription that combines the conventional
self-energies with so-called pinch parts from vertex and box diagrams in such
a manner that the modified self-energies are gauge independent and exhibit 
desirable theoretical properties \cite{cor,deg}.
We calculate the pinch part $\Delta A(s)$ in $R_\xi$ gauge by means of the
$S$-matrix PT framework elaborated in Ref.~\cite{deg}.
We choose the elastic scattering of two fermions via a Higgs boson in the $s$
channel as the reference process.
Our result is independent of this choice \cite{cor,deg}.
The relevant Feynman diagrams are depicted in Fig.~\ref{fig:two}.
In the formulation of Ref.~\cite{deg}, the corresponding amplitudes reflecting
the interactions of the vector bosons with the external fermions are described
in terms of matrix elements of Fourier transforms of time-ordered products of
current operators.
Through successive current contractions with the longitudinal four-momenta
found in the propagators and vertices of the massive vector bosons, Ward
identities are triggered, after which the relevant pinch contributions are
identified with amplitudes involving appropriate equal-time commutators of
currents.
Setting aside the details, the pinch contribution is found to be
\begin{eqnarray} 
\Delta A(s)&=&\frac{G}{\pi}(s-M^2)
\left\{\frac{1}{2}\left[A_0\left(M_W^2\right)-A_0\left(\xi_WM_W^2\right)
\right]\right.
\nonumber\\
&&{}+\left[\frac{1}{4}(s+M^2)+M_W^2\right]B_0\left(s,M_W^2,M_W^2\right)
\nonumber\\
&&{}-\left.\frac{1}{4}(s+M^2)B_0\left(s,\xi_WM_W^2,\xi_WM_W^2\right)
+\frac{1}{2}(W\to Z)\right\}.
\label{eq:hpt}
\end{eqnarray}
The second and third lines of Eq.~(\ref{eq:hpt}) agree with Eq.~(2.21) of
Ref.~\cite{pap}, where the seagull and tadpole diagrams were omitted because
they were not needed for the purpose of that paper.
As expected, the PT self-energy of the Higgs boson,
\begin{equation}
a(s)=A(s)+\Delta A(s),
\label{eq:hpi}
\end{equation}
is independent of $\xi_W$ and $\xi_Z$ for all values of $s$.
Furthermore, $\Delta A(s)$ vanishes at $s=M^2$, so that the one-loop
expression for $M$ and the tree-level one for $\Gamma$ are not affected by the
application of the PT.

The one-loop radiative corrections to physical observables characterizing the
production or decay of a real Higgs boson involve its wave-function
renormalization constant $Z$.
The on-shell definition of the latter, given in Eq.~(\ref{eq:zos}), contains
the term $\re A^\prime(M^2)$, which is the source of the threshold
singularities at $M=2M_V$.
To see that, let us consider the expression
\begin{equation}
\left.\frac{\partial}{\partial s}B_0\left(s,M_V^2,M_V^2\right)
\right|_{s=M^2}=\left\{
\begin{array}{ll}
-\frac{\displaystyle 1}{\displaystyle M^2}
\left(1+\frac{\displaystyle {\cal A}}{\displaystyle \sqrt{1-{\cal A}}}
\arsinh\sqrt{-\frac{\displaystyle 1}{\displaystyle {\cal A}}}\right),
& \quad {\cal A}<0, \\
& \\
-\frac{\displaystyle 1}{\displaystyle M^2}
\left[1+\frac{\displaystyle {\cal A}}{\displaystyle \sqrt{1-{\cal A}}}
\left(\arcosh\sqrt{\frac{\displaystyle 1}{\displaystyle {\cal A}}}
-i\frac{\displaystyle \pi}{\displaystyle 2}\right)\right],
& \quad 0<{\cal A}<1, \\
& \\
-\frac{\displaystyle 1}{\displaystyle M^2}
\left(1-\frac{\displaystyle {\cal A}}{\displaystyle \sqrt{{\cal A}-1}}
\arcsin\sqrt{\frac{\displaystyle 1}{\displaystyle {\cal A}}}\right),
& \quad {\cal A}>1,
\end{array}\right.
\label{eq:bp1}
\end{equation}
where ${\cal A}=4M_V^2/M^2$,\footnote{In Eq.~(\ref{eq:bp1}), we also consider
the case ${\cal A}<0$, which provides a convenient starting point for the
analytic continuation to be performed in Section~\ref{sec:uns}.}
which appears in $A^\prime(M^2)$ with the prefactor
$-(G/\pi)\left(M^4/4\right.$\break $\left.-M^2M_V^2+3M_V^4\right)$.
Equation~(\ref{eq:bp1}) follows from \cite{pr}
\begin{equation}
B_0\left(s,M_V^2,M_V^2\right)=\frac{1}{\epsilon}-\gamma_E+\ln
\frac{4\pi\mu^2}{M_V^2}+2-2\sqrt{1-\frac{4\left(M_V^2-i\varepsilon\right)}{s}}
\arsinh\sqrt{-\frac{s}{4\left(M_V^2-i\varepsilon\right)}}+O(\epsilon),
\end{equation}
where $\gamma_E$ is Euler's constant.
As $M$ approaches $2M_V$ from below, Eq.~(\ref{eq:bp1}) develops a real 
singularity proportional to $(2M_V-M)^{-1/2}$.
In a one-loop calculation, one expands $Z$ as $Z=1+\re A^\prime(M^2)+O(g^4)$,
which then exhibits the same threshold singularity.
As $M$ surpasses $2M_V$, this threshold singularity is shifted from the real
part to the imaginary one and, therefore, it does not affect $Z$.
Since the prefactors of $B_0\left(s,\xi_VM_V^2,\xi_VM_V^2\right)$ in 
Eq.~(\ref{eq:hxi}) vanish at $s=M^2$, there are no threshold singularities at
the unphysical thresholds $M=2\sqrt{\xi_V}M_V$ in $R_\xi$ gauge.
On the other hand, in the PT framework, there are no gauge-dependent
thresholds, and Eq.~(\ref{eq:bp1}) appears in $a^\prime(M^2)$ with the same
prefactor as in $A^\prime(M^2)$.
In conclusion, the threshold singularities are gauge independent and only
affect physical thresholds.

\section{Pole formulation
\label{sec:pol}}

We now explain how the threshold singularities are avoided in the pole
scheme.
Throughout this section, we assume that the threshold particles are stable.
The general case of unstable threshold particles will be discussed in
Section~\ref{sec:uns}.
As is well known \cite{sir,kni}, the on-shell and pole definitions of mass and
widths, given in Eqs.~(\ref{eq:mos}), (\ref{eq:gos}), (\ref{eq:mpo}), and
(\ref{eq:gpo}), are equivalent through next-to-leading order, i.e.\ through
$O(g^2)$ and $O(g^4)$, respectively.
In particular, this implies that the perturbative expansion of $m_2\Gamma_2$
through $O(g^4)$ resembles the one of $M\Gamma$,
\begin{equation}
M\Gamma=-\im A^{(1)}(M^2)\left[1+\re A^{(1)\prime}(M^2)\right]
-\im A^{(2)}(M^2)+O(g^6),
\label{eq:os1}
\end{equation}
where the superscripts refer to the number of quantum loops, and thus also
suffers from threshold singularities.
In fact, Taylor expanding the right-hand side of Eq.~(\ref{eq:gpo}) about
$m_2$ and retaining only terms through $O(g^4)$, we obtain
\begin{eqnarray}
m_2\Gamma_2&=&-\im A^{(1)}\left(m_2^2\right)
+m_2\Gamma_2\re A^{(1)\prime}\left(m_2^2\right)
-\im A^{(2)}\left(m_2^2\right)+O(g^6)
\nonumber\\
&=&-\im A^{(1)}\left(m_2^2\right)\left[1+\re A^{(1)\prime}\left(m_2^2\right)
\right]-\im A^{(2)}\left(m_2^2\right)+O(g^6),
\label{eq:po1}
\end{eqnarray}
which is equivalent to Eq.~(\ref{eq:os1}) through $O(g^4)$ and thus also prone
to threshold singularities.
In order to avoid the latter, we have to undo the Taylor expansion in 
Eq.~(\ref{eq:po1}), i.e.\ we have to substitute
\begin{eqnarray}
\re A^{(1)\prime}\left(m_2^2\right)&=&
\frac{\im A^{(1)}\left(m_2^2\right)-\im A^{(1)}\left(\bar s\right)}
{m_2\Gamma_2}
+O(g^4)
\nonumber\\
&=&-1
-\frac{\im A^{(1)}\left(m_2^2-im_2\Gamma_2^{(0)}\right)}{m_2\Gamma_2^{(0)}}
+O(g^4),
\label{eq:sub}
\end{eqnarray}
where, consistent with our approximation, we have replaced $\Gamma_2$ with the
tree-level width $\Gamma_2^{(0)}=-\im A^{(1)}\left(m_2^2\right)/m_2$.
By the same token, the substitution rule of Eq.~(\ref{eq:sub}) allows us to
eliminate, in the spirit of the pole scheme, the threshold singularities
which have been encountered in the on-shell scheme
\cite{fle,bff,bvv,hww,hff,pr,jeg}.
To that end, we abandon Eq.~(\ref{eq:bp1}) and instead substitute
\begin{equation}
\left.\re\frac{\partial}{\partial s}B_0\left(s,m_{2,V}^2,m_{2,V}^2\right)
\right|_{s=m_2^2}=\frac{\im B_0\left(m_2^2,m_{2,V}^2,m_{2,V}^2\right)
-\im B_0\left(\bar s,m_{2,V}^2,m_{2,V}^2\right)}{m_2\Gamma_2}+O(g^2),
\label{eq:bp2}
\end{equation}
where $m_{2,V}$ is the pole mass of the threshold particles, together with
\begin{eqnarray}
\im B_0\left(m_2^2,m_{2,V}^2,m_{2,V}^2\right)&=&\pi\sqrt{1-a}\theta(1-a),
\label{eq:im1}\\
\im B_0\left(\bar s,m_{2,V}^2,m_{2,V}^2\right)&=&
f(a,\gamma)+2\pi\sqrt{1-a}\theta(1-a)\theta(\gamma),
\label{eq:im2}
\end{eqnarray}
where $a=4m_{2,V}^2/m_2^2$ and $\gamma=\Gamma_2/m_2$.
Here, we have used the auxiliary function
\begin{eqnarray}
f(a,\gamma)&\equiv&-2\im\left(\sqrt{1-\bar a}
\arsinh\sqrt{-\frac{1}{\bar a}}\right)
\nonumber\\
&=&\sign(\gamma)
\frac{\sqrt2}{b}\left\{\frac{1}{2}\sqrt{b(c-b)+a}\right.
\nonumber\\
&&{}\times\ln\left[\frac{1}{a}\left(b+c+\sqrt{(b-1)(c+a-1)}
+\sqrt{(b+1)(c-a+1)}\right)\right]
\nonumber\\
&&{}-\left.\sqrt{b(c+b)-a}
\arctan\frac{\sqrt{b+1}+\sqrt{c-a+1}}{\sqrt{b-1}+\sqrt{c+a-1}}\right\},
\label{eq:f}
\end{eqnarray}
where $\bar a=4m_{2,V}^2/\bar s$, $b=\sqrt{1+\gamma^2}$,
$c=\sqrt{(a-1)^2+\gamma^2}$, and $\sign(x)=\theta(x)-\theta(-x)$.
The term proportional to $\theta(1-a)$ in Eq.~(\ref{eq:im2}) guarantees that
Eqs.~(\ref{eq:im1}) and (\ref{eq:im2}) refer to the same Riemann sheet, so
that their difference appearing in Eq.~(\ref{eq:bp2}) vanishes in the limit
$\Gamma_2\to0$ and the derivative expression on the left-hand side of that
equation is recovered if $m_2\ne2m_{2,V}$.
The factors $\theta(\gamma)$ and $\sign(\gamma)$ in Eqs.~(\ref{eq:im2}) and
(\ref{eq:f}), respectively, make sure that the case $\gamma<0$ is covered,
too, a generalization that will be useful in Section~\ref{sec:uns}.
At threshold, where $m_2=2m_{2,V}$, Eq.~(\ref{eq:bp2}) leads to
\begin{equation}
\left.\re\frac{\partial}{\partial s}B_0\left(s,m_{2,V}^2,m_{2,V}^2\right)
\right|_{s=m_2^2}
=\frac{1}{m_2^2}\left[\frac{\pi}{\sqrt{2\gamma}}\left(1+\frac{\gamma}{2}
\right)-2+O\left(\gamma^{3/2}\right)\right]+O(g^2),
\end{equation}
i.e.\ the threshold singularity of the on-shell scheme is automatically
regularized in the pole scheme by the width $\Gamma_2$ of the primary
particle.

Finally, we explain how Eq.~(\ref{eq:sub}) is generalized to higher orders.
For that purpose, we rewrite Eq.~(\ref{eq:gpo}) as
\begin{equation}
m_2\Gamma_2=-\frac{\im A\left(m_2^2\right)}
{1-\left[\im A\left(m_2^2\right)-\im A\left(\bar s\right)\right]/
(m_2\Gamma_2)}.
\label{eq:ide}
\end{equation}
Solving Eq.~(\ref{eq:ide}) for $m_2\Gamma_2$, we recover Eq.~(\ref{eq:gpo}), 
so that the two expressions are equivalent.
The usefulness of Eq.~(\ref{eq:ide}) may be appreciated by observing that, in
order to calculate $\Gamma_2$ to $O(g^{2n+2})$, we only need to insert the
$O(g^{2n})$ expression for $\Gamma_2$ on the right-hand side of that equation.
Comparing Eq.~(\ref{eq:ide}) with Eq.~(\ref{eq:gos}), we see that
\begin{equation}
Z_2=\frac{1}
{1-\left[\im A\left(m_2^2\right)-\im A\left(\bar s\right)\right]/
(m_2\Gamma_2)}
\label{eq:zpo}
\end{equation}
plays the r\^ole of the wave-function renormalization constant for unstable 
particles in the pole scheme.
This is to be compared with its counterpart in the on-shell scheme, given in
Eq.~(\ref{eq:zos}).

\section{Unstable threshold particles
\label{sec:uns}}

So far, we have assumed the threshold particles to be stable.
We now allow for them to have a finite pole width $\Gamma_{2,V}$.
This can be achieved by replacing in the loop amplitude $A(m_2^2)$ the square 
of their mass $m_{2,V}^2$ by the complex position
$\bar s_V=m_{2,V}^2-im_{2,V}\Gamma_{2,V}$ of the pole of their propagator.
In this way, the substitution rule given in Eq.~(\ref{eq:bp2}) becomes
\begin{equation}
\left.\re\frac{\partial}{\partial s}B_0\left(s,\bar s_V,\bar s_V\right)
\right|_{s=m_2^2}=\frac{\im B_0\left(m_2^2,\bar s_V,\bar s_V\right)
-\im B_0\left(\bar s,\bar s_V,\bar s_V\right)}{m_2\Gamma_2}+O(g^2).
\label{eq:bp3}
\end{equation}
The expression for $\im B_0\left(\bar s,\bar s_V,\bar s_V\right)$ has a
structure analogous to Eq.~(\ref{eq:im2}).
Comparing
$4\bar s_V/\bar s=\left[a\left(1+\gamma_V^2\right)/(1+\gamma\gamma_V)\right]/
\left[1-i(\gamma-\gamma_V)/(1+\gamma\gamma_V)\right]$, where
$\gamma_V=\Gamma_{2,V}/m_{2,V}$, with $\bar a=a/(1-i\gamma)$, we see that in
Eq.~(\ref{eq:f}) $a$ and $\gamma$ are to be replaced by
$a\left(1+\gamma_V^2\right)/(1+\gamma\gamma_V)$ and 
$(\gamma-\gamma_V)/(1+\gamma\gamma_V)$, respectively.
This leads to
\begin{equation}
\im B_0\left(\bar s,\bar s_V,\bar s_V\right)
=f\left(a\frac{1+\gamma_V^2}{1+\gamma\gamma_V},
\frac{\gamma-\gamma_V}{1+\gamma\gamma_V}\right)
+2\pi\sqrt{1-a}\theta(1-a)\theta(\gamma-\gamma_V).
\label{eq:im3}
\end{equation}
Setting $\gamma=0$ in Eq.~(\ref{eq:im3}), we obtain
\begin{equation}
\im B_0\left(m_2^2,\bar s_V,\bar s_V\right)
=f\left(a\right(1+\gamma_V^2\left),-\gamma_V\right).
\label{eq:im4}
\end{equation}
The second term on the right-hand side of Eq.~(\ref{eq:im3}) ensures that
Eqs.~(\ref{eq:im3}) and (\ref{eq:im4}) refer to the same Riemann sheet, so
that Eq.~(\ref{eq:bp3}) reduces to Eq.~(\ref{eq:bp2}) in the limit
$\Gamma_{2,V}\to0$.

In a situation when the threshold particles are much more unstable than the
primary particle, i.e.\ if $\Gamma_2/m_2\ll\Gamma_{2,V}/m_{2,V}$, we may take
the limit $\Gamma_2\to0$ on the right-hand side of Eq.~(\ref{eq:bp3}) and thus
effectively return to the derivative expression on the left-hand side of that
equation, which reads
\begin{eqnarray}
\left.\re\frac{\partial}{\partial s}B_0\left(s,\bar s_V,\bar s_V\right)
\right|_{s=m_2^2}&=&-\frac{1}{m_2^2}\re\left(1+
\frac{\bar a_V}{\sqrt{1-\bar a_V}}\arsinh\sqrt{-\frac{1}{\bar a_V}}\right)
\nonumber\\
&=&-\frac{1}{m_2^2}\left\{1+\frac{a}{\sqrt2c_V}\left\{\frac{1}{2}
\left(\sqrt{c_V-a+1}-\gamma_V\sqrt{c_V+a-1}\right)\right.\right.
\nonumber\\
&&{}\times
\ln\left[\frac{1}{ab_V^2}\left(b_V+b_Vc_V
+\sqrt{(b_V-1)\left(b_Vc_V+ab_V^2-1\right)}\right.\right.
\nonumber\\
&&{}+\left.\left.\sqrt{(b_V+1)\left(b_Vc_V-ab_V^2+1\right)}\right)\right]
\nonumber\\
&&{}-\left(\sqrt{c_V+a-1}+\gamma_V\sqrt{c_V-a+1}\right)
\nonumber\\
&&{}\times\left.\left.
\arctan\frac{\sqrt{b_V+1}+\sqrt{b_Vc_V-ab_V^2+1}}
{\sqrt{b_V-1}+\sqrt{b_Vc_V+ab_V^2-1}}\right\}\right\},
\label{eq:bp4}
\end{eqnarray}
where $\bar a_V=4\bar s_V/m_2^2$, $b_V=\sqrt{1+\gamma_V^2}$, and
$c_V=\sqrt{(a-1)^2+a^2\gamma_V^2}$.
At threshold, we have
\begin{equation}
\left.\re\frac{\partial}{\partial s}B_0\left(s,\bar s_V,\bar s_V\right)
\right|_{s=m_2^2}
=\frac{1}{m_2^2}\left[\frac{\pi}{2\sqrt{2\gamma_V}}(1+\gamma_V)-2
+O\left(\gamma_V^2\right)\right]+O(g^2),
\end{equation}
i.e.\ the threshold singularity is now entirely regularized by the width
$\Gamma_{2,V}$ of the threshold particles.

\section{Discussion
\label{sec:dis}}

We are now in a position to investigate the phenomenological implications of
our results.
The SM Higgs boson with mass $m_2$ in the vicinity of $2m_{2,Z}$ dominantly
decays to a pair of $W$ bosons, with a branching fraction of about 90\%
\cite{pr}.
Therefore, we choose the partial width of this decay as an example to 
illustrate the threshold singularity and its removal.

The complete one-loop radiative correction to this observable was obtained
within the on-shell scheme in Refs.~\cite{fle,bvv,hww}.
Its structure is exhibited in Eq.~(\ref{eq:os1}) if we include in
$\im A^{(1)}(M^2)$ and $\im A^{(2)}(M^2)$ only intermediate states containing
a $W^+W^-$ pair.
Specifically, $-\im A^{(1)}(M^2)$ represents the tree-level result, written
with $G_\mu$, $Z=1+\re A^{(1)\prime}(M^2)$ is the Higgs-boson wave-function
renormalization constant of Eq.~(\ref{eq:zos}), and $-\im A^{(2)}(M^2)$
comprises the proper $HW^+W^-$ vertex correction, the $HW^+W^-$ coupling and
$W$-boson wave-function renormalization constants, and the real-photon
bremsstrahlung correction.
As we have seen in Section~\ref{sec:os}, the Taylor expansion of 
Eq.~(\ref{eq:gpo}), given in Eq.~(\ref{eq:po1}), is equivalent to 
Eq.~(\ref{eq:os1}).
In the following, we work in the pole scheme, on the basis of
Eq.~(\ref{eq:po1}).
All the ingredients of Eq.~(\ref{eq:po1}) may be found, in analytic form, in
Ref.~\cite{hww}.

As explained in the context of Eq.~(\ref{eq:bp1}), the threshold singularities
arise from the term
$\left.\re\partial B_0\left(s,m_{2,V}^2,m_{2,V}^2\right)/\partial s
\right|_{s=m_2^2}$, which is contained in
$\re A^{(1)\prime}\left(m_2^2\right)$.
If the threshold particles are stable, these threshold singularities are
regularized by $\Gamma_2$ according to the substitution rule of 
Eq.~(\ref{eq:bp2}).
In the case of unstable threshold particles, this substitution rule is
generalized by including $\Gamma_{2,V}$ as described in Eq.~(\ref{eq:bp3}).
In the limiting case $\Gamma_2/m_2\ll\Gamma_{2,V}/m_{2,V}$, we recover the 
derivative expression of Eq.~(\ref{eq:bp4}), in which the threshold 
singularity is regularized by $\Gamma_{2,V}$.
In the case under consideration, we have $V=Z$ and
$(\Gamma_2/m_2):(\Gamma_{2,V}/m_{2,V})\approx1:7$, so that the second method
of regularization, which incorporates both $\Gamma_2$ and $\Gamma_{2,V}$,
should be most appropriate, while the third one, which is solely based on
$\Gamma_{2,V}$, should provide a good approximation.
On the other hand, the first scheme, which only includes $\Gamma_2$, should be
unrealistic in this particular case.

In our numerical analysis, we use $m_{2,W}=80.391$~GeV, $m_{2,Z}=91.153$~GeV,
and $\Gamma_{2,Z}=2.493$~GeV \cite{ewwg}, and adopt the residual input
parameters from Ref.~\cite{pdg}.
We remind the reader that $m_2^2=m_1^2/\left(1+\Gamma_1^2/m_1^2\right)$ and
$\Gamma_2^2=\Gamma_1^2/\left(1+\Gamma_1^2/m_1^2\right)$, where $m_1$ and 
$\Gamma_1$ can be identified with the measured values \cite{sir}.
We evaluate the Higgs-boson total decay width $\Gamma_2$, which enters
Eqs.~(\ref{eq:bp2}) and (\ref{eq:bp3}), in the Born approximation.
In Fig.~\ref{fig:three}, we show the $H\to W^+W^-$ partial decay width at one
loop in the pole scheme as a function of $m_2$ in the vicinity of the
threshold at $m_2=2m_{2,Z}$.
The evaluation from the Taylor-expanded expression given in
Eq.~(\ref{eq:po1}) on the basis of Eq.~(\ref{eq:bp1}) (dotted line), which
exhibits a threshold singularity, is compared with the one where this
threshold singularity is jointly regularized by $\Gamma_2$ and $\Gamma_{2,Z}$
according to the substitution rule of Eq.~(\ref{eq:bp3}) (solid line).
For comparison, we also display the tree-level result (dashed line).
We observe that the regularized result smoothly interpolates across the 
threshold region and merges with the unregularized result sufficiently far
away from the threshold.
Below (above) threshold, the regularization leads to an increase (decrease) of
the result relative to the unregularized case.
In the threshold region, the regularized correction increases the Born result
by about 7\%.
In Fig.~\ref{fig:four}, we compare the regularized result shown in
Fig.~\ref{fig:three} (solid line) with the results based on the
regularizations by $\Gamma_2$ according to Eq.~(\ref{eq:bp2}) (dashed line)
and by $\Gamma_{2,Z}$ according to Eq.~(\ref{eq:bp4}) (dot-dashed line).
For reference, we also show the unregularized result (dotted line).
Comparing the dashed and solid lines, we observe that $\Gamma_{2,Z}$ plays a
crucial r\^ole in the regularization of the threshold singularity, as
anticipated above.
On the other hand, we infer from the closeness of the dot-dashed and solid
lines that the relative importance of $\Gamma_2$ in the combined
regularization approach is minor, due to the smallness of $\Gamma_2/m_2$ as
compared to $\Gamma_{2,V}/m_{2,V}$.

\section{Conclusions
\label{sec:con}}

As explained in the text, the threshold singularity, associated with the
conventional on-shell wave-function renormalization constant, affects the
production and decay rates of the unstable particle whenever its mass is
degenerate with the sum of masses of an interacting pair of virtual particles
that form an $S$-wave state.
It is important to realize that, since the wave-function renormalization 
constant is a universal prefactor in all decay and production amplitudes, the
associated singularity affects all production and decay processes of the
unstable particle.

\begin{figure}[ht]
\begin{center}
\centerline{\epsfig{figure=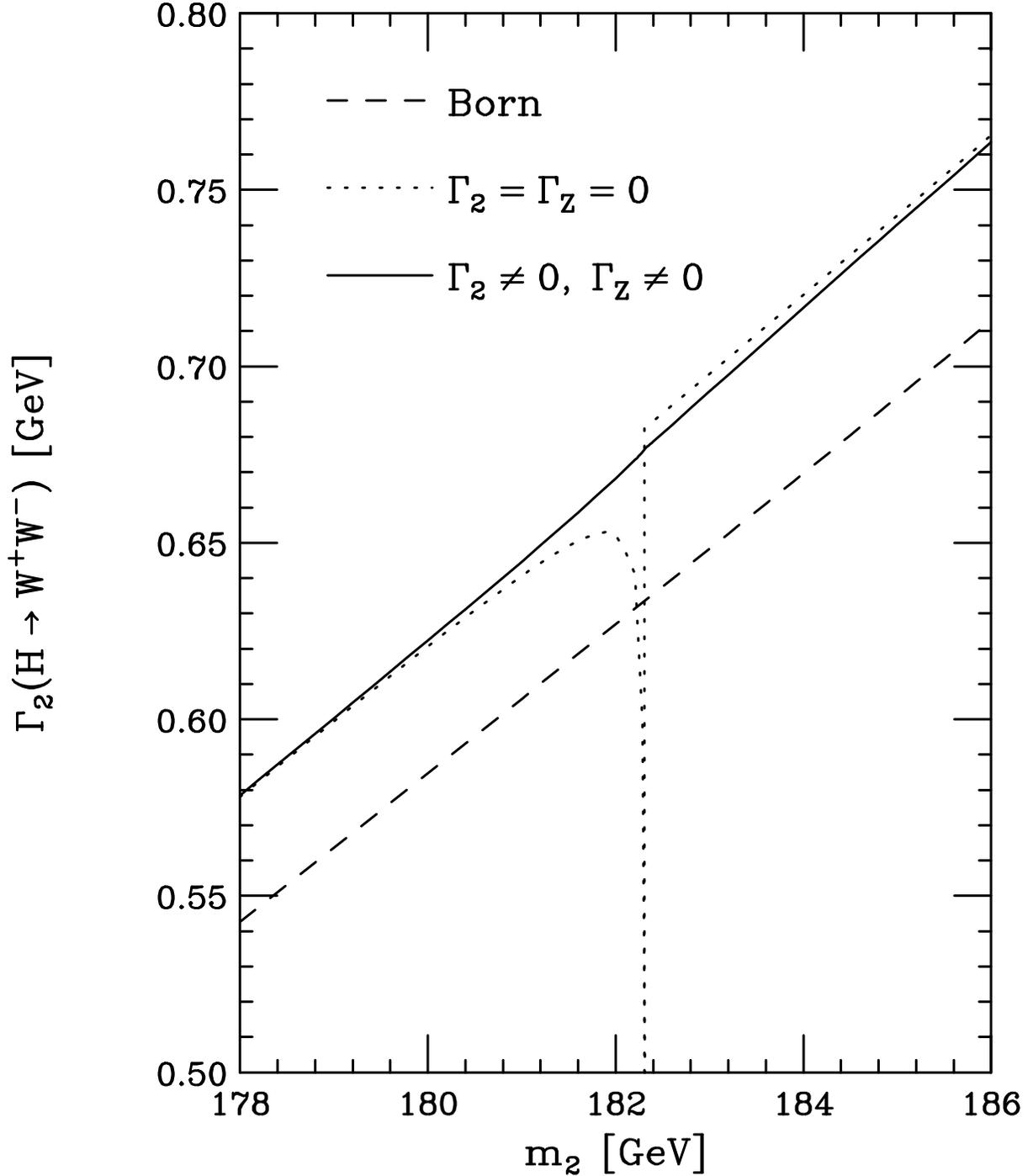,width=16cm}}
\caption{$H\to W^+W^-$ partial decay width at one loop in the pole scheme as a
function of the Higgs-boson pole mass $m_2$ in the vicinity of the threshold 
at $m_2=2m_{2,Z}$.
The evaluation from the Taylor-expanded expression given in
Eq.~(\ref{eq:po1}) on the basis of Eq.~(\ref{eq:bp1}) (dotted line), which
exhibits a threshold singularity, is compared with the one where this
threshold singularity is jointly regularized by $\Gamma_2$ and $\Gamma_{2,Z}$
according to the substitution rule of Eq.~(\ref{eq:bp3}) (solid line).
For comparison, also the tree-level result is shown (dashed line).}
\label{fig:three}
\end{center}
\end{figure}

\begin{figure}[ht]
\begin{center}
\centerline{\epsfig{figure=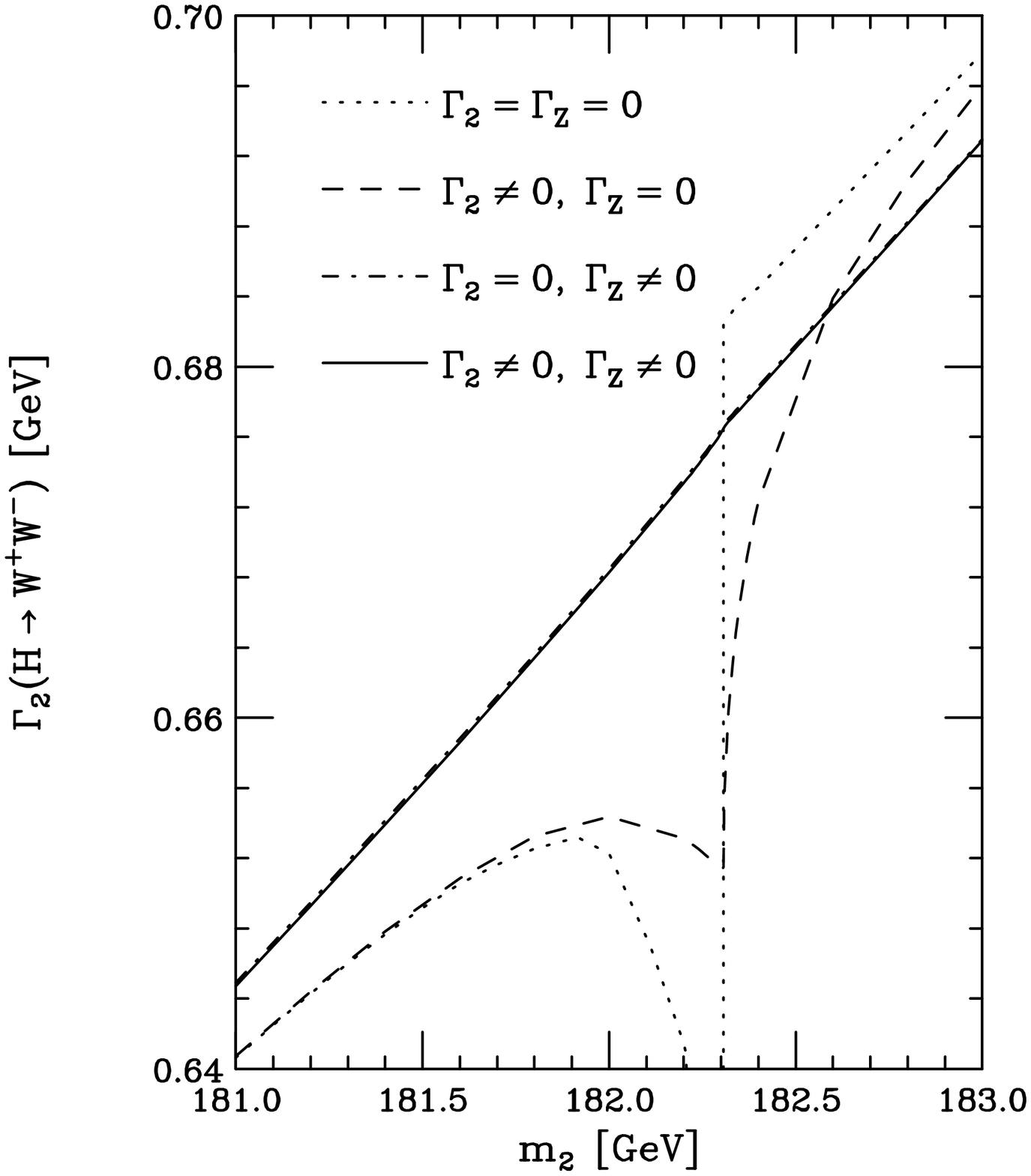,width=16cm}}
\caption{$H\to W^+W^-$ partial decay width at one loop in the pole scheme as a
function of the Higgs-boson pole mass $m_2$ in the vicinity of the threshold 
at $m_2=2m_{2,Z}$.
The threshold singularity is (a) not regularized (dotted line) or regularized
(b) by $\Gamma_2$ according to Eq.~(\ref{eq:bp2}) (dashed line), (c) by
$\Gamma_{2,Z}$ according to Eq.~(\ref{eq:bp4}) (dot-dashed line), or (d) by
$\Gamma_2$ and $\Gamma_{2,Z}$ according to Eq.~(\ref{eq:bp3}) (solid line).}
\label{fig:four}
\end{center}
\end{figure}

For definiteness, we focussed our analysis on a case of phenomenological
interest, namely the $H\to W^+W^-$ decay of the SM Higgs boson when its mass
$M$ is very close to $2M_Z$.
By examining the one-loop Higgs-boson self-energy in $R_\xi$ gauge, we showed
that the threshold singularity is gauge independent and, for that reason, also
affects the conventional wave-function renormalization constant in the PT
framework.
We then demonstrated how the one-loop threshold singularity is removed in the
pole formulation.
In particular, the conventional on shell wave-function renormalization
constant given in Eq.~(\ref{eq:zos}) is replaced by Eq.~(\ref{eq:zpo}), which
plays the r\^ole of the wave-function renormalization for the unstable
particle in the pole scheme.
When the virtual particles in the mass-degenerate pair are themselves
unstable, their widths can also be used to tame the threshold singularity.
We then showed how the two regularizing effects, associated with the widths of
the primary and threshold particles, can be combined in a unified analysis.
The various effects are illustrated for the $H\to W^+W^-$ decay width in
Figs.~\ref{fig:three} and \ref{fig:four}. 
As a byproduct, we presented in Eqs.~(\ref{eq:hxi}) and (\ref{eq:hpi}) the
complete Higgs-boson self-energies at one loop in $R_\xi$ gauge and in the
PT framework, respectively.

\vspace{1cm}
\begin{center}
{\bf Acknowledgements}
\end{center}
\smallskip

B.A.K. thanks Oleg Yakovlev for fruitful discussions concerning 
Ref.~\cite{mel}.
A.S. is grateful to the Theory Group of the 2$^{\rm nd}$ Institute for
Theoretical Physics for the hospitality extended to him during a visit when
this manuscript was prepared.
The work of B.A.K. was supported in part by the Deutsche
Forschungsgemeinschaft through Grant No.\ KN~365/1-1, by the
Bundesministerium f\"ur Bildung und Forschung through Grant No.\ 05~HT9GUA~3,
and by the European Commission through the Research Training Network
{\it Quantum Chromodynamics and the Deep Structure of Elementary Particles}
under Contract No.\ ERBFMRX-CT98-0194.
The work of C.P.P. was supported by the German Academic Exchange Service
(DAAD) through Grant No.\ A/97/00746.
The work of A.S. was supported in part by the Alexander von Humboldt
Foundation through Research Award No.\ IV~USA~1051120~USS, and by the
National Science Foundation through Grant No.\ PHY-9722083.

\newpage

\newpage

\newpage

\newpage

\end{document}